\pdfoutput=1

\documentclass{PoS}

\usepackage{amsmath}
\usepackage{xspace}
\usepackage{listings}

\usepackage{siunitx}

\newcommand{\MCatNLO}{M\protect\scalebox{0.8}{C}@N\protect\scalebox{0.8}{LO}\xspace}

\newcommand{\LO}{L\scalebox{0.8}{O}\xspace}

\newcommand{\NLO}{N\scalebox{0.8}{LO}\xspace}
\newcommand{\NNLO}{N\scalebox{0.8}{NLO}\xspace}

\newcommand{\Madgraph}{M\protect\scalebox{0.8}{AD}G\protect\scalebox{0.8}{RAPH}\xspace}

\newcommand{\aMCatNLO}{aM\protect\scalebox{0.8}{C}@N\protect\scalebox{0.8}{LO}\xspace}

\newcommand{\Rivet}{R\protect\scalebox{0.8}{IVET}\xspace}

\newcommand{\HepMC}{\protect\texttt{HepMC}\xspace}
\newcommand{\HepMCWeightContainer}{\protect\texttt{HepMC::WeightContainer}\xspace}
\newcommand{\aMCfast}{\protect\scalebox{0.8}{A}MC\protect\scalebox{0.8}{FAST}\xspace}

\newcommand{\Sherpa}{S\protect\scalebox{0.8}{HERPA}\xspace}
\newcommand{\fastNLO}{\protect\scalebox{0.8}{FAST}NLO\xspace}
\newcommand{\APPLgrid}{APPL\protect\scalebox{0.8}{GRID}\xspace}
\newcommand{\MCgrid}{MC\protect\scalebox{0.8}{GRID}\xspace}

\long\def\symbolfootnote[#1]#2{\begingroup%
\def\thefootnote{\fnsymbol{footnote}}\footnote[#1]{#2}\endgroup}

\newcommand{\mr}[1]{\mathrm{#1}}

\newcommand{\bea}{\begin{eqnarray}}
\newcommand{\eea}{\end{eqnarray}}
\newcommand{\bi}{\begin{itemize}}
\newcommand{\ei}{\end{itemize}}

\newcommand{\obs}{\langle O\rangle}
\newcommand{\alphaS}{\alpha_s}
\newcommand{\BME}{\mathrm{B}}

\newcommand{\VIME}{\mathrm{VI}}
\newcommand{\VIMEp}{\mathrm{VI}^\prime}

\newcommand{\KPME}{\mathrm{KP}}

\newcommand{\RME}{\mathrm{R}}

\newcommand{\DSME}{\mathrm{D}_S}

\newcommand{\ntrial}{n_\text{trial}}
\newcommand{\Ntrial}{N_\text{trial}}

\newcommand{\muR}{\mu_R}
\newcommand{\muF}{\mu_F}
\newcommand{\muRt}{\tilde\mu_R}
\newcommand{\muFt}{\tilde\mu_F}

\newcommand{\lR}{l_R}
\newcommand{\lF}{l_F}
\newcommand{\cR}[1]{c_R^{(#1)}}
\newcommand{\cF}[2]{c_{F,#1}^{(#2)}}
\newcommand{\cFp}[2]{c_{F,#1}^{\,\prime\,(#2)}}
\newcommand{\cFpp}[2]{c_{F,#1}^{\,\prime\prime\,(#2)}}

\title{Fast evaluation of theoretical uncertainties with Sherpa and MCgrid}

\ShortTitle{Fast evaluation of theoretical uncertainties with Sherpa and MCgrid}

\author{\hfill ZU-TH 21/15, MCnet-15-18}

\author{\speaker{Enrico Bothmann}\\  
        II. Physikalisches Institut, Georg-August-Universit\"at G\"ottingen, 37077 G\"ottingen, Germany\\
        E-mail: \email{enrico.bothmann@phys.uni-goettingen.de}}

\author{Marek Sch\"onherr\\
        Institut f\"ur Theoretische Physik, Universit\"at Z\"urich, 8057 Z\"urich, Switzerland\\
        E-mail: \email{marek.schoenherr@physik.uzh.ch}}

\author{Steffen Schumann\\
        II. Physikalisches Institut, Georg-August-Universit\"at G\"ottingen, 37077 G\"ottingen, Germany\\
        E-mail: \email{steffen.schumann@phys.uni-goettingen.de}}

\abstract{The determination of theoretical error estimates and PDF/$\alpha_s$-fits require 
fast evaluations of differential cross sections for varied QCD input parameters.
These include PDFs, the strong coupling constant $\alpha_S$
and the renormalisation and factorisation scales.
Beyond leading order QCD, a full dedicated calculation for each set of parameters
is often too time-consuming, certainly when performing PDF-fits.
We report on two methods to overcome this issue for any QCD NLO calculation:
The novel event-reweighting feature in \textsc{Sherpa} and the automated generation of 
interpolation grids using the recently introduced \textsc{MCgrid} interface.
For \textsc{MCgrid} we present the newly added support for \textsc{fastNLO} tables
and highlight some future developments.}

\FullConference{XXIII International Workshop on Deep-Inelastic Scattering\\
		April 27--May 1, 2015\\
		Dallas, Texas}

\begin{document}

\section{Introduction}
In the last decade basically the complete automation of QCD \NLO calculations 
has been achieved \cite{Carli:2015qta}. The predictions of these calculations 
depend on the QCD input parameters, such as the renormalisation scale $\mu_\mathrm{R}$,
the factorisation scale $\mu_\mathrm{F}$,
the strong coupling $\alphaS(\mu_\mathrm{R}^2)$
and the parton density functions (PDFs) $f_a(x, \mu_\mathrm{F}^2)$
for the gluon and the (anti-)quarks.

These input parameters need to be varied when we want to calculate theory uncertainties,
e.g.\ when we follow the convention to vary the scales with factors 0.5 and 2
to estimate the effect of higher-order corrections.
But for QCD cross section calculations beyond the \LO it usually takes days
to achieve a precision that is comparable to data of current experiments, e.g.\
at the Large Hadron Collider.
So a full re-evaluation for a given set of input parameter values uses up lots of CPU cycles.
This problem is even more severe for another application of parameter variation:
Fits of $\alphaS$ and PDFs,
where many re-evaluations are needed before the fit converges.
Certainly for PDFs, which have many degrees of freedom in a fit,
performing dedicated calculations for each PDF candidate value becomes impossible.

The solution is to identify how a prediction depends on the QCD input parameters.
Then we can reuse the independent and CPU-intense parts for subsequent re-evaluations.
This has been achieved for those types of QCD calculations,
where the input parameters (or a known function of them)
enter the calculation as prefactors to the identified independent parts~\cite{Bern:2013zja}.

There are two approaches to re-evaluate these independent parts
for another choice of input parameter values.
One is direct reweighting:
We present here a novel feature of the 2.2.0 release of the Monte-Carlo event generator \Sherpa~\cite{Gleisberg:2008ta}
for performing an on-the-fly reweighting.
Another approach is to create an interpolation grid,
which allows for even faster re-evaluations.
Currently there are two implementations for those grids,
\APPLgrid~\cite{Carli:2010rw} and \fastNLO~\cite{Kluge:2006xs},
and two interfaces for their automated creation using NLO events,
\aMCfast~\cite{Bertone:2014zva} and \MCgrid~\cite{DelDebbio:2013kxa}.
Here we discuss the new \MCgrid 2.0 release and present its newly added support
for \fastNLO and for filling the fixed-order expansion of \MCatNLO calculations.
We also present a method
for quantifying the dependency of a parton shower algorithm on the QCD input parameters.
This is a first step towards the goal of correctly taking care of these dependencies
in a reweighting or an interpolation grid creation.

\section{QCD NLO Events} \label{secEvents}

Consider the structure of an \NLO QCD calculation
using Catani-Seymour subtraction~\cite{Catani:1996vz},
which consists of the following building blocks:
The Born term ($\mr{B}$), the 
virtual correction plus $\mr{I}$-operator of the integrated subtraction terms 
($\mr{VI}$), the $\mr{K}$- and $\mr{P}$-operators of the integrated subtraction 
terms including the collinear factorisation remainders ($\mr{KP}$), and the 
real-emission ($\mr{R}$) and correlated subtraction ($\DSME$) terms:
\begin{equation}\label{eq:nlo}
  \begin{split}
    \obs^\text{\NLO}
    \,=\;&\lim\limits_{N\to\infty}\,\frac{1}{\Ntrial}\;
    \Bigg\{
      \sum_i^{N_B}\Bigg\{
        \BME(\Phi_{B,i})
        +\VIME(\Phi_{B,i})
        +\KPME(\Phi_{B, i},x_{a/b}^\prime)
      \Bigg\}\;O(\Phi_{B,i})\\
         &\hspace*{24mm}
          {}+\sum_i^{N_R} \left[\vphantom{\sum_j}
          \RME(\Phi_{R_i})\;O(\Phi_{R_i})
          -\sum_j\DSME(\Phi_{B,j,i}\cdot\Phi_{R_i|B}^j)\;O(\Phi_{B,j,i})
        \right]
   \Bigg\}\,,
  \end{split}
\end{equation}
with $\Ntrial=\sum_i^N\left.\ntrial\right._i$ the sum of the number of trials for each event
and with the observable cuts and projections $O(\Phi)$, where $\Phi$ gives the phase space point.
The $O(\Phi)$ are only dependent on the final state and so are independent of the QCD input parameters.
We will now make explicit the dependency structure wich respect to the input parameters of each of the building blocks $\mr{B}$, $\mr{VI}$, $\mr{KP}$ and $\DSME$.
The simplest are the ``Born-like'' terms $X=\BME,\,\RME,\,\DSME$,
where a combination of QCD input parameters occurs as a total prefactor,
as is well-known from the \LO factorisation theorem:
\begin{equation}
    X(\Phi) = f_a(x_a,\muF^2)\;f_b(x_b,\muF^2)\;
    \alphaS^{n+p}(\muR^2)\; X'(\Phi)\,,
\end{equation}
where $n$ is the number of strong vertices in the LO diagrams and $p$ is either 0 ($\BME$) or 1 ($\RME, \DSME$).
Here and in the following the primed quantity ($X'$) on the right has no dependency on the QCD input parameters.
We call it the weight of the corresponding term ($X$).
As one can see, for the evaluation of the PDFs the longitudinal 
momentum fractions $x_{a/b}$ have to be known in addition to the weight.
The structure of the $\VIME$ term is more complicated due to the occurring renormalisation scale logarithms:
\begin{equation}\label{eq:NLO-scale-var}
    \VIME(\Phi_B) = f_a(x_a,\muF^2)\;f_b(x_b,\muF^2)\;
    \bigg[\alphaS^{n+1}(\muR)\;
    \VIMEp(\Phi_B)
          +\left(\cR{0}\lR+\tfrac{1}{2}\,\cR{1}\lR^2\right)
    \bigg]\,,
\end{equation}
with coefficients $\cR{i}$ and scale logarithms $\lR=\log\frac{\muR^2}{\muRt^2}$,
where $\muRt$ is the scale at which the $\cR{i}$ had been evaluated in the first place.
So for re-evaluating the VI term,
not only the weight $\VIMEp$
and the momentum fractions $x_{a/b}$ need to be provided,
but also the coefficients $\cR{i}$ and the original scale $\muRt$.
The last missing building block are the KP terms:
\begin{equation}
  \begin{split}
    \KPME(\Phi_B,x_{a/b}^\prime)\,=\;&\alphaS^{n+1}(\muR)\,
    \bigg[
      f_b(x_b,\muF^2)
      \left(
        f_a^q \cF{a}{0} + f_a^q(x_a^\prime) \cF{a}{1}
        +f_a^g \cF{a}{2} + f_a^g(x_a^\prime) \cF{a}{3}
      \right)
    \\
         &\hspace*{20mm}{}+
      f_a(x_a,\muF^2)
      \left(
        f_b^q \cF{b}{0} + f_b^q(x_b^\prime) \cF{b}{1}
        +f_b^g \cF{b}{2} + f_b^g(x_b^\prime) \cF{b}{3}
      \right)
    \bigg]\,,
  \end{split}
\end{equation}
with $x_{a/b}^\prime$ being the ratio of $x_{a/b}$ before and after the initial-state branching,
and the coefficients
\begin{equation}
\cF{a/b}{i}=\cFp{a/b}{i}+\cFpp{a/b}{i}\,\lF\quad \text{for} \quad
i=0,\ldots,3 \quad \text{with} \quad \lF=\log\frac{\muF^2}{\muFt^2}.
\end{equation}
The $c_F^{\,\prime}$ and $c_F^{\,\prime\prime}$ are independent of the QCD input parameters.
Again, $\tilde{\mu}_{F}$ is the scale at which the $c_F$ had been evaluated.
The functions $f^{q/g}$ give the correct combination of PDFs
for the possible initial-state branchings, depending on whether a quark or a gluon is split,
cf.~\cite{DelDebbio:2013kxa}.

All of these weights, coefficients and kinematics information can be passed
in a standard \HepMC~\cite{Dobbs:2001ck} event record:
The record comes with a \HepMCWeightContainer object,
which can be used to store all weights and coefficients.
This must of course be supported by the event generator,
as done by e.g.\ \Sherpa~\cite{Gleisberg:2008ta}.

With this information provided, the weights and coefficients can be combined 
with any arbitrary choice of QCD input parameters.
The same is true for the \NLO expansion of \MCatNLO and (partly) for multi-leg merging
calculations, but their structure will not be discussed here to keep this note short.
It is not true however for parton showers, their inherent all-orders resummation 
structure renders such a direct decomposition to prefactors composed of input parameters
and a fixed number of weights/coefficients impossible.
This lack of a correct treatment for variations in the shower
results in a caveat for PDF variations of multi-leg merging calculations:
They are not fully exact, because for them the PDFs enter via the shower branching 
probabilities determining the rejection weights for emissions already at the matrix 
element level.

\section{Sherpa-Internal Reweighting} \label{secReweighting}
\begin{figure}[hbtp]
  \centering
  \includegraphics[width=0.7\textwidth]{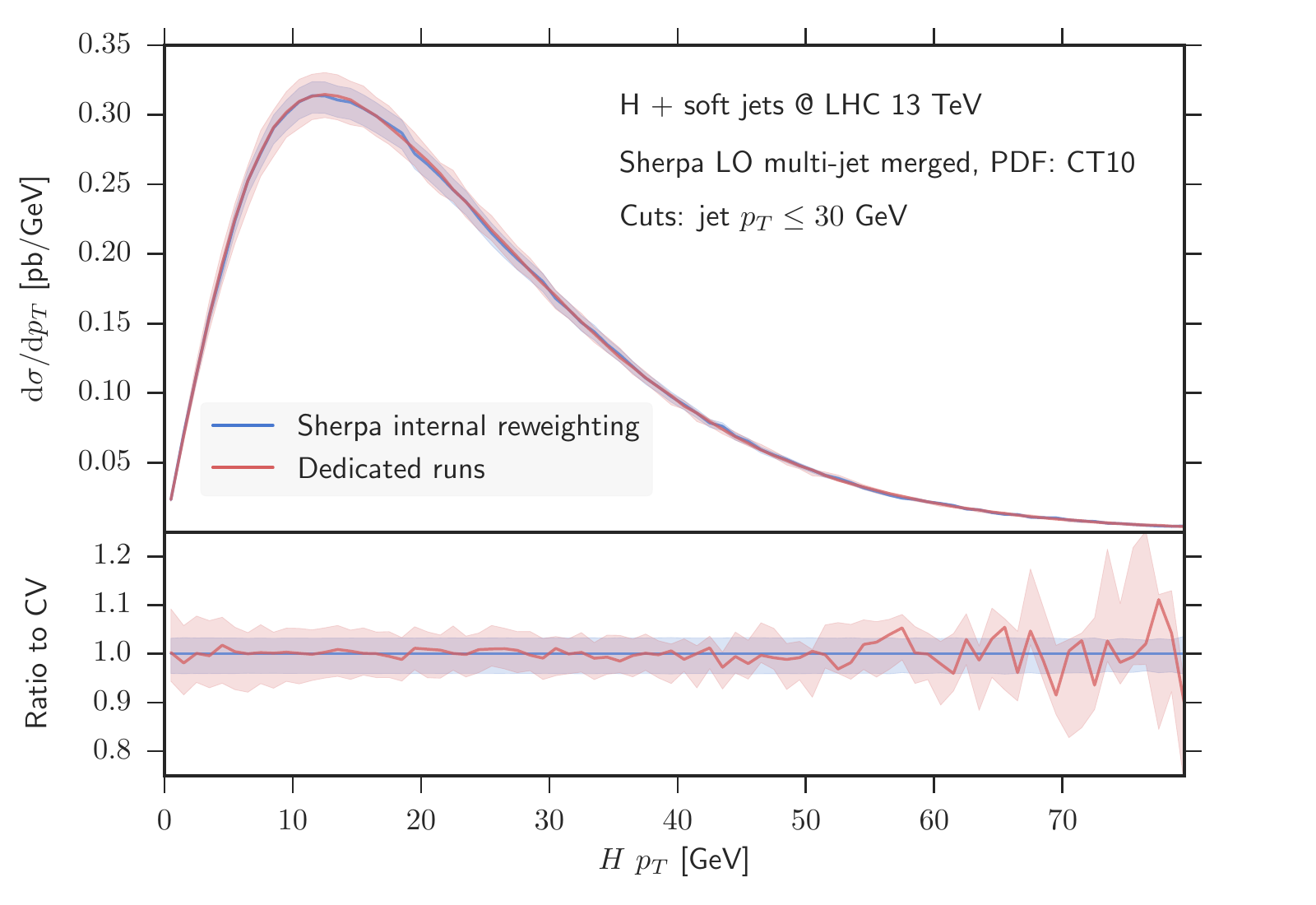}
  \caption{Validation for the new internal reweighting feature of \Sherpa 2.2.0. A PDF band for the CT10 set~\cite{Lai:2010vv} is generated using a single run with reweighting enabled to generate all PDF error eigenvalues at the same time. This is compared with the band for the same PDF set, but now doing a dedicated event generation for each set, i.e.\ without any reweighting.} 
  \label{figReweightingValidation}
\end{figure}

The newly added internal reweighting feature of \Sherpa 2.2.0
uses the extraction of weights and coefficients sketched in the previous 
section to perform an on-the-fly reweighting for variations of the  QCD input 
parameters. Any number of combinations of scales and PDFs can be specified 
prior to the run. The results are inserted into the \HepMCWeightContainer 
object and thus can be either saved to disk, programmatically accessed or 
directly passed via the internal interface to the \Rivet analysis 
framework~\cite{Buckley:2010ar}. \LO, \NLO, the fixed-order expansion of 
\MCatNLO and/or multi-jet merged runs are supported. The parton shower can 
be enabled in the run, but its dependency on the QCD input parameters can 
not be resolved yet as discussed in the previous section. The reweighting 
is then incomplete and the resulting prediction is not exact.
Figure~\ref{figReweightingValidation} illustrates the validation for the 
internal reweighting approach. Shown is the PDF variation plot for the
transverse-momentum distribution in Higgs-boson production at the LHC,
comparing via a single run with reweighting, with dedicated event generation 
runs for the modified PDF sets. For this example, the reweighting run 
takes approximately 4 times longer than a usual run, but generates the 
same information (with less statistical fluctuation) as 52 dedicated runs.
In this example (and all the following ones) the input PDF has been interfaced
using LHAPDF~6~\cite{Buckley:2014ana}.


\section{Automated Grid Creation using MCgrid} \label{secMCgrid}
The internal reweighting feature of \Sherpa is useful for evaluating theory uncertainties for an observable.
But when speed becomes very critical,
because the number of re-evaluations needed becomes large (e.g.\ when fitting PDFs),
more specialised methods must be used to reduce CPU time drastically.
One established approach
is based on cross section interpolation grids that represent
a given differential cross section binned in the incoming partonic momentum fractions $x_{a/b}$
and the process scale $\mu_R^2=\mu_F^2=Q^2$.
There are two implementations for those grids
called \APPLgrid~\cite{Carli:2010rw} and \fastNLO~\cite{Kluge:2006xs}.
While interpolating grids are defined only for a specific final state observable,
they allow for a very quick re-evaluation of the respective cross section
and have considerably reduced disk space requirements,
compared to full event data.
Until recently, interfaces to fill these grids were only available for process-specialised code.
This has changed, as two interfaces to multi-purpose Monte Carlo event generators
are now available.

One is called \aMCfast~\cite{Bertone:2014zva}:
It interfaces \LO and \NLO predictions
using the FKS scheme~\cite{Frederix:2009yq} generated with
\Madgraph{}5\_\aMCatNLO~\cite{Alwall:2011uj} to \APPLgrid.

\begin{figure}[hbtp]
  \centering
  \includegraphics[width=0.7\textwidth]{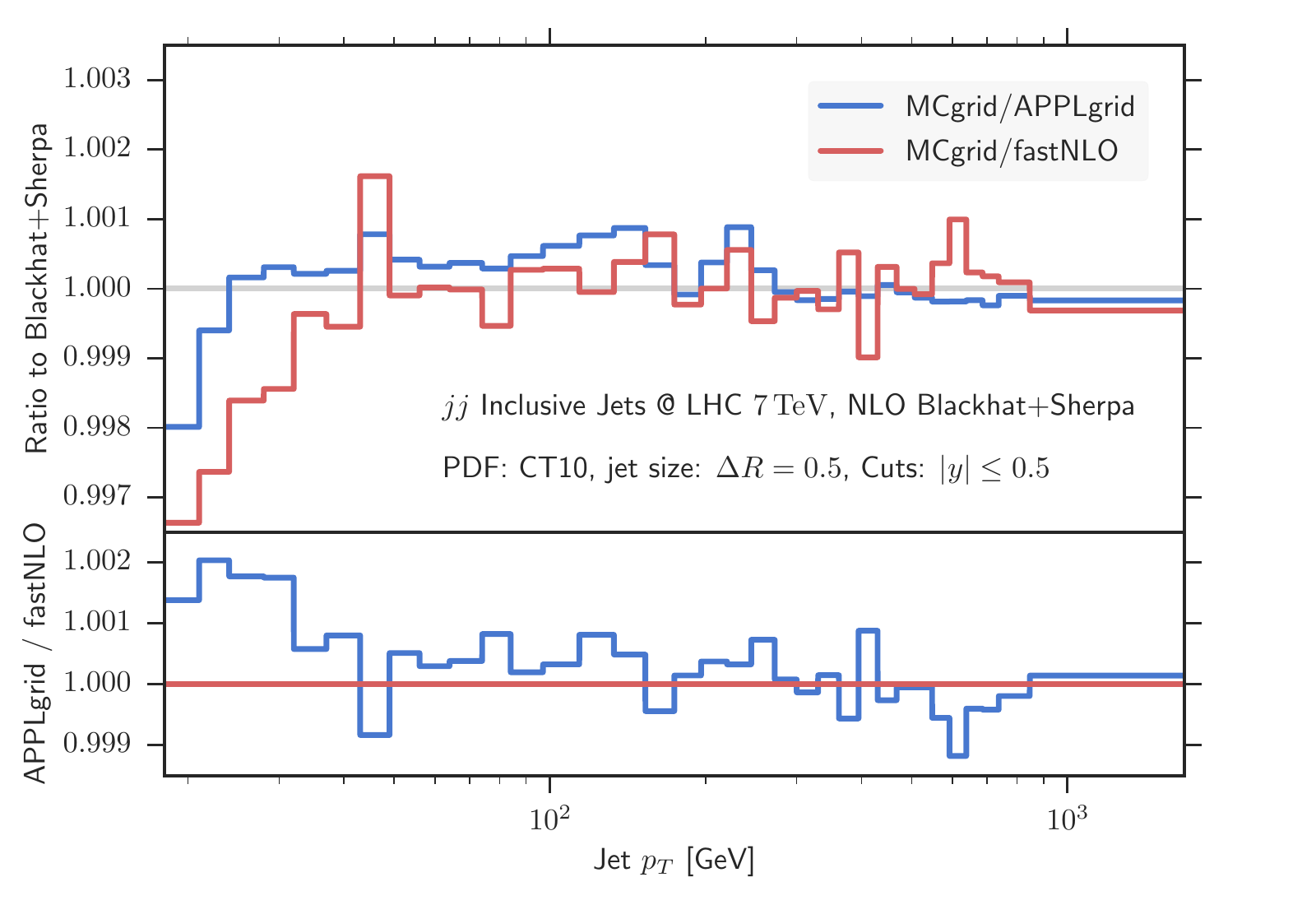}
  \caption{Validation for newly added support for \fastNLO grids in the 2.0 release of \MCgrid. An inclusive jet $p_T$ distribution is generated using a single run. At the same time, interpolation grids for \fastNLO and \APPLgrid are written out via \MCgrid. Their convolution with the same QCD input parameters used in the run are compared with the results from the direct run. The deviation is of a per-mille order. This has been achieved without any optimisations of the grid architecture (number of nodes/interpolation order).}
  \label{figMCgridValidation}
\end{figure}

The second one is called \MCgrid, which is implemented as a plugin to \Rivet~\cite{Buckley:2010ar},
such that (new or existing) \Rivet analyses can be modified to create and fill interpolations grids.
\MCgrid reads in the events in the \HepMC~\cite{Dobbs:2001ck} event data format,
such that in principle every Monte Carlo event generator
can be used to produce grids via \MCgrid.
However, all required event weights must be inserted into the weight vector of the \HepMC event.
Currently, only the format that \Sherpa writes into the weight vector is supported by \MCgrid,
which is based on the Catani-Seymour dipole subtraction scheme as illustrated in sec.~\ref{secEvents}.
As for \aMCfast, since the first release of \MCgrid~\cite{DelDebbio:2013kxa},
\APPLgrid interpolation grids can be filled with \LO and \NLO events.
Here we report on two newly added features in the 2.0 release of \MCgrid:
the support of \fastNLO interpolation grids
and the exact treatment of the fixed-order expansion of \MCatNLO events.
This is the first time the \fastNLO package can be interfaced
to a multi-purpose Monte Carlo event generator.
With \APPLgrid and \fastNLO, all currently available interpolation tools
for fixed-order QCD cross sections can now be used in conjunction with MCgrid.
For more details on the revisions made in \MCgrid 2.0, see~\cite{Bothmann:2015}.
Figure~\ref{figMCgridValidation} illustrates for the case of inclusive jet production 
at LHC energies, that \fastNLO and \APPLgrid interpolation grids created with 
\MCgrid both reproduce the distribution of the observable with per-mille level accuracy,
much smaller than experimental and inherent theoretical errors.

\section{Application: Residual PDF Dependencies of the Parton Shower} \label{secPartonShowers}

As has been explored earlier~\cite{Gieseke:2004tc},
varying the QCD input parameters in parton showers
can have a sizeable effect on the result.
The possibility to correctly vary the full fixed-order expansion of an \MCatNLO calculation%
---via the internal reweighting of \Sherpa or using interpolation grids created with \MCgrid---%
allows for the quantification of the residual dependencies of the parton showers
that are beyond the fixed-order approximation.
Focusing on the PDF being used, the method would go as follows: First, generate PDF error bands for a given observable for two different PDF sets A and B. Use the event generation for the central value member of A to obtain distributions for all members of B via internal reweighting or an interpolation grid. Now the reweighted/interpolated PDF band can be compared to the directly generated band for B.
Because the PDF has been varied everywhere but in the parton shower, any (statistically significant) deviation in this comparison between the two B bands would indicate the associated observable bin's sensitivity to the PDF 
input of the parton shower.
Finding sizeable deviations can motivate an implementation
for tracking the dependencies of the parton shower.
The sensitive bins would then serve as test cases to check if such an implementation works as expected,
as the deviations must vanish then
(or shrink if an approximation is used as e.g.\ only tracking the first parton shower emission).
Interpolation grids filled with additional information for the parton shower
would help in the fitting of PDFs to data that are not appropriately described by fixed-order calculations.

\begin{figure}[hbtp]
  \centering
  \includegraphics[width=0.7\textwidth]{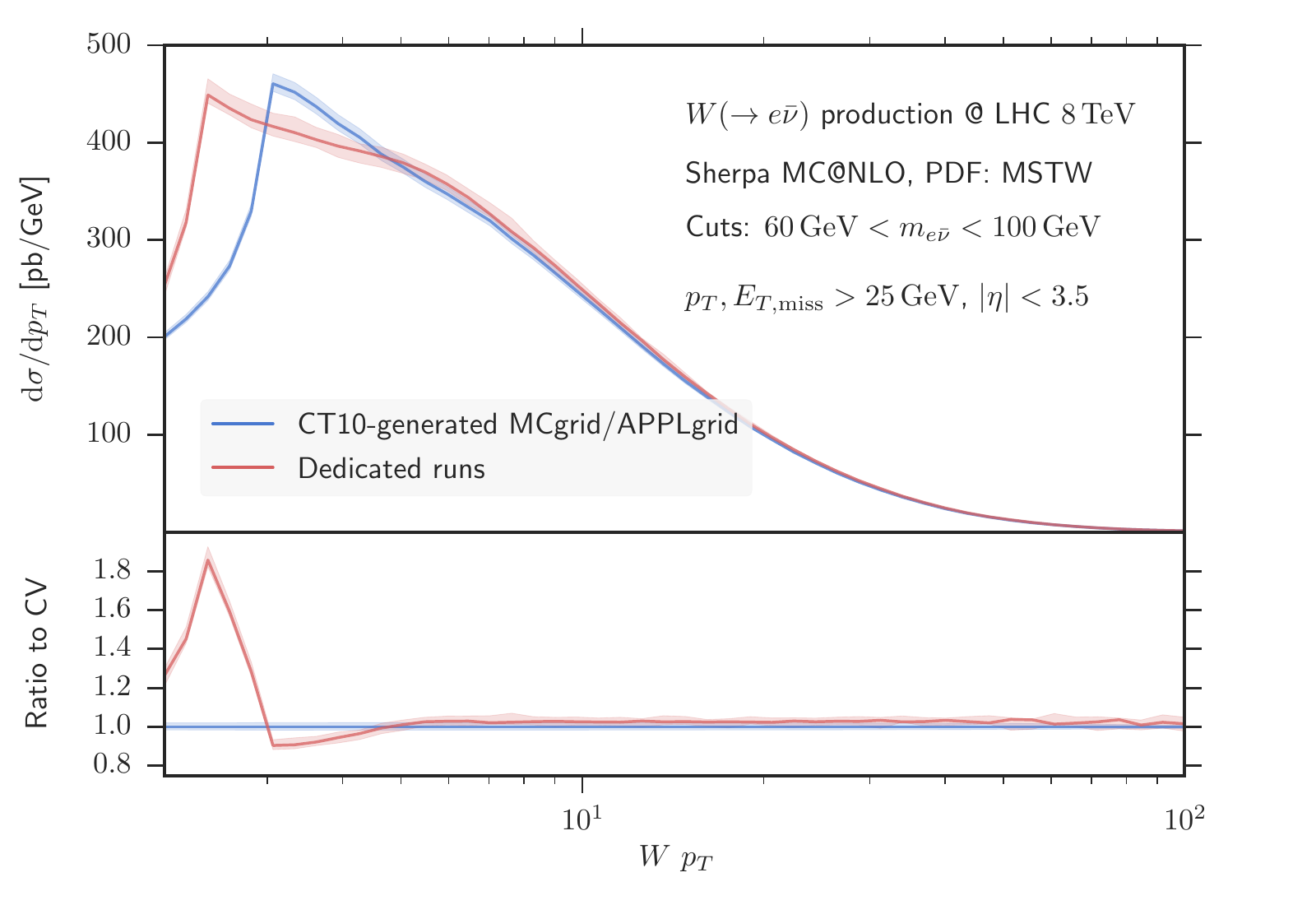}
  \caption{A Study of the residual PDF dependence of the parton shower of a grid that has been produced in an \MCatNLO event generation with the central set of CT10 PDF. The MSTW-convolution of the grid deviates from the dedicated MSTW event generation in the low $p_T \lesssim \SI{5}{GeV}$ region.}
  \label{figResidualPDFDependency}
\end{figure}

In fig.~\ref{figResidualPDFDependency} we exemplify such a study
with $W$ boson production at the LHC, considering the $p_T$ distribution of the boson.
The grid has been filled with residual dependencies on the PDF set that has been used 
during its creation: the CT10~\cite{Lai:2010vv} central value member.
When convoluting it with the MSTW pdf set~\cite{Martin:2009iq}
and comparing it with a set of dedicated runs,
we observe deviations in the low $p_T \lesssim \SI{5}{GeV}$ region.
This corresponds to a shift of the Sudakov peak position.
The peak's shape and position is sensitive to multiple emissions
and to the behaviour of the PDFs at very small scales.
Therefore a careful consideration of these sensitivities
must precede a definite conclusion.

\section{Conclusion}

We have presented two novel methods for fast re-evaluations of fixed-order cross sections suitable 
for the determination of theoretical uncertainties or PDF/$\alpha_s$-fits in an automated way.
Although both methods are based on reusing all the parts of the calculation that are not dependent on QCD input parameters,
they do not require event data to be written out.

The internal reweighting feature of \Sherpa 2.2.0 is doing the reweighting on-the-fly, i.e. on an event-per-event basis.
This is especially useful for simultaneously generating all distributions needed
for specifying the theory errors for a given observable.
\MCgrid on the other hand is an interface for automated grid production using general-purpose event generators and \Rivet.
Because interpolation grids cut down the CPU time needed for a re-evaluation to the order of milliseconds, they are apt for PDF fitting.

The support of both methods for filling the fixed-order expansion of an \MCatNLO calculation
gives rise to a way to study the dependencies of the parton shower on the QCD input parameters.
We have exemplified this for the transverse momentum distribution of the $W$ boson.

In the future we plan to extend these studies,
aiming to eventually implementing a method for taking into account
(approximate) parton shower dependencies
for the internal reweighting of \Sherpa
and for the creation of interpolation grids via \MCgrid.
Other features we want to add to \MCgrid are \NNLO predictions
and the support for multi-jet merging calculations.

\section*{Acknowledgments}

We acknowledge financial support from BMBF under contract 05H12MG5, the 
Swiss National Foundation (SNF) under contract PP00P2--128552 and 
from the EU MCnetITN research network funded under Framework Programme 
7 contract PITN-GA-2012-315877.

\end{document}